\documentclass[journal]{IEEEtran}
\IEEEoverridecommandlockouts

\ifCLASSINFOpdf
\else
\fi

\usepackage{graphicx}
\usepackage{graphics}
\usepackage{xcolor}
\usepackage[hidelinks]{hyperref}
\usepackage{amsmath,amsfonts}
\usepackage{cleveref}
\Crefformat{figure}{#2Fig.~#1#3}
\Crefmultiformat{figure}{Figs.~#2#1#3}{ and~#2#1#3}{, #2#1#3}{ and~#2#1#3}
\usepackage[safe]{tipa}
\usepackage{tipx}
\usepackage{scalerel}
\usepackage{textcomp}
\usepackage{threeparttable}
\usepackage{tablefootnote}

\usepackage{algpseudocode} 
\usepackage{algorithm} 
\usepackage{url}
\usepackage{multirow}
\usepackage{multicol}
\usepackage{soul}
\usepackage[normalem]{ulem}
\usepackage{comment}
\usepackage{diagbox}
\usepackage{amssymb}

\usepackage{array}
\usepackage{acronym}
\usepackage[caption=false,font=footnotesize]{subfig}
\usepackage{stfloats}
\usepackage{verbatim}
\usepackage{multirow}
\usepackage{adjustbox}
\usepackage[utf8]{inputenc}
\usepackage[mathscr]{euscript}
\usepackage{amsmath,amssymb}

\usepackage{booktabs}

\def\BibTeX{{\rm B\kern-.05em{\sc i\kern-.025em b}\kern-.08em
    T\kern-.1667em\lower.7ex\hbox{E}\kern-.125emX}}
\usepackage{balance}

\usepackage[backend=biber,,citestyle=numeric-comp,bibstyle=ieee,sorting=none,minbibnames=1,maxbibnames=6,doi=false,url=false,eprint=false]{biblatex}
\newacro{CIM}{compute-in-memory}
\newacro{ADC}{analog-to-digital converter}
\newacro{SAR}{successive approximation register}
\newacro{DAC}{digital-to-analog converter}
\newacro{DDRF}{direct digital-to-RF}
\newacro{BB}{baseband}
\newacro{UBB}{ultra-broadband}
\newacro{RAN}{radio access network}
\newacro{SCPA}{switched-capacitor power amplifier}
\newacro{DTC}{digital-to-time converter}
\newacro{RF-DAC}{radio-frequency digital-to-analog converter}
\newacro{IF}{intermediate frequency}
\newacro{TTD}{true-time-delay}
\newacro{DDMW}{direct digital-to-mm-wave}
\newacro{TX}{transmitter}
\newacro{DTX}{digital \ac{TX}}
\newacro{RX}{receiver}
\newacro{TRX}{transceiver}
\newacro{SNR}{signal-to-noise ratio}
\newacro{BS}{basestation}
\newacro{UE}{user equipment}
\newacro{AF}{array factor}
\newacro{TIPA}{two-input power amplifier}
\newacro{RF}{radio frequency}
\newacro{RAN}{radio access network}
\newacro{QoS}{quality of service}
\newacro{V2V}{vehicle-to-vehicle}
\newacro{OOB}{out-of-band}
\newacro{TAF}{time-approximation filter}
\newacro{LUT}{look-up table}
\newacro{OSR}{over-sampling rate}
\newacro{BER}{bit error rate}
\newacro{PAA}{phased-array antenna}
\newacro{AoA}{angle of arrival}
\newacro{AoD}{angle of departure}
\newacro{ML}{machine learning}
\newacro{AI}{artificial intelligence}
\newacro{FoM}{figure of merit}
\newacro{TI}{time-interleaved}
\newacro{NN}{neural network}
\newacro{CNN}{convolutional \ac{NN}}
\newacro{LoS}{line-of-sight}
\newacro{PVT}{process, voltage and temperature}
\newacro{AR}{augmented reality}
\newacro{FDD}{frequency division duplexing}
\newacro{DNN}{deep \ac{NN}}
\newacro{LSTM}{long short-term memory}
\newacro{OAM}{orbital angular momentum}
\newacro{WDM}{wavelength division multiplexing}
\newacro{IoT}{internet-of-things}
\newacro{CMOS}{complementary metal-oxide-semiconductor}
\newacro{EDA}{electronic design automation}
\newacro{ATE}{automated test equipment}
\newacro{SDR}{software defined radio}
\newacro{FPGA}{field programmable gate array}
\newacro{LNA}{low noise amplifier}
\newacro{PA}{power amplifier}
\newacro{PCM}{phase-change material}
\newacro{CR}{cognitive radio}
\newacro{DPD}{digital pre-distortion}
\newacro{SHM}{sub-harmonic mixer}
\newacro{TIA}{trans-impedance amplifier}
\newacro{LDO}{low-dropout regulator}
\newacro{APT}{average power tracking}
\newacro{ET}{envelope tracking}
\newacro{EER}{envelope elimination and restoration}
\newacro{GaN}{gallium nitride}
\newacro{RFIC}{\ac{RF} integrated circuits}
\newacro{GaAs}{gallium arsenide}
\newacro{DBS}{deep brain stimulation}
\newacro{RF}{radio frequency}
\newacro{TMS}{trans-cranial magnetic stimulation}
\newacro{tDCS}{trans-cranial direct-current stimulation}
\newacro{TES}{trans-cranial electrical stimulation}
\newacro{COTS}{commercial, off-the-shelf}
\newacro{SAR}{specific absorption rate}
\newacro{PCB}{printed circuit board}
\newacro{EMG}{electromyography}
\newacro{SoC}{system-on-a-chip}
\newacro{EVM}{error vector magnitude}
\newacro{MIMO}{multiple-input, multiple-output}
\newacro{QAM}{quadrature amplitude modulation}
\newacro{OFDM}{orthogonal frequency-division multiplexing}
\newacro{SC}{single-carrier}
\newacro{FR}{frequency range}
\newacro{FR1}{frequency range~1}
\newacro{FR2}{frequency range~2}
\newacro{FR3}{frequency range~3}
\newacro{DFT}{discrete Fourier transform}
\newacro{SFDR}{spurious-free dynamic range}
\newacro{SNDR}{signal-to-noise and distortion ratio}
\newacro{PAPR}{peak-to-average power ratio}
\newacro{CIC}{cascaded integrator-comb}
\newacro{PSD}{power spectral density}
\newacro{LVT}{low-threshold-voltage}
\newacro{ULVT}{ultra-low-threshold-voltage}
\newacro{ZOH}{zero-order hold}

\begin{document}

\title{A Hybrid Digital/Analog Polyphase Filtering Approach for High-Speed RF-DACs}
\author{Behdad Jamadi,~\IEEEmembership{Student Member,~IEEE,}
        Jungmin Lee,~\IEEEmembership{Student Member,~IEEE,}
        Tarun Cousik,~\IEEEmembership{Member,~IEEE,}
        Rohit Rangaraj,~\IEEEmembership{Student Member,~IEEE,}
        Jeffrey S. Walling,~\IEEEmembership{Senior Member,~IEEE}\\
\thanks{All authors are with the Bradley Department of Electrical and Computer Engineering, Virginia Tech, Blacksburg, VA 24061 USA. (Corresponding author: Jeffrey S. Walling, e-mail: jswalling@vt.edu). This work was supported by the National Science Foundation (NSF) under grant \#2314813. Simulation code and models available at \url{https://github.com/noyades/Non-uniform-Sampling-RF-DAC}.}}


\maketitle

\begin{abstract}
This paper proposes a hybrid polyphase interpolation for RF-DACs that removes the need for full-rate digital commutation by using passive combining on the RF-DAC itself. A $P$-branch polyphase FIR filter is factored into $M$ streams, each formed by a $(P/M)$-way digital commutator at $f_\mathrm{RF}/M$ and driving a dedicated RF-DAC slice, giving either partial digital ($1<M<P$) or fully analog ($M=P$) recombination. We show that the slice array acts as a single full-rate DAC with an $M$-times longer zero-order hold, derive yield laws for slice gain, timing and unit-capacitor mismatch, and find that with retiming the binding requirement is sub-picosecond matching of the $f_\mathrm{RF}$ clock arrival, so the best choice is the smallest $M$ that closes digital timing. Two 28-nm digital macros, designed for a $12.8$\,GSps output rate and verified by post-layout simulation, illustrate a 4-phase variant that closes timing up to $14.4$\,GSps with its digital logic at $f_\mathrm{RF}/4$ and an 8-phase variant that closes timing up to $16$\,GSps with its digital logic at $f_\mathrm{RF}/8$. The analog recombination is evaluated with a behavioral model of the slice array. The architectures target wideband \ac{SDR} platforms, including cognitive radio, electronic warfare, and satellite communications.
\end{abstract}

\section{Introduction}

Wideband \acfp{SDR} are central elements for modern wireless communications systems, including those in commercial and military usage. They enable frequency-agile transceivers that can accommodate a wide variety of waveforms from single hardware platforms~\cite{mitola1995software}. \acp{SDR} have become relatively commonplace below $7$\,GHz (\ac{FR1}) \cite{watanabe2021_5g}, owing to the increasing sampling rate that is available for data converters. The advent of the \ac{RF-DAC} is a primary enabler for wireless \acp{TX}, in that it can directly synthesize \ac{BB} waveforms and output them directly at \ac{RF}~\cite{luschas2004_rfdac,yoo2011scpa,alavi2014idac}.

\begin{figure}[t]
  \centering
  \includegraphics[width=3in]{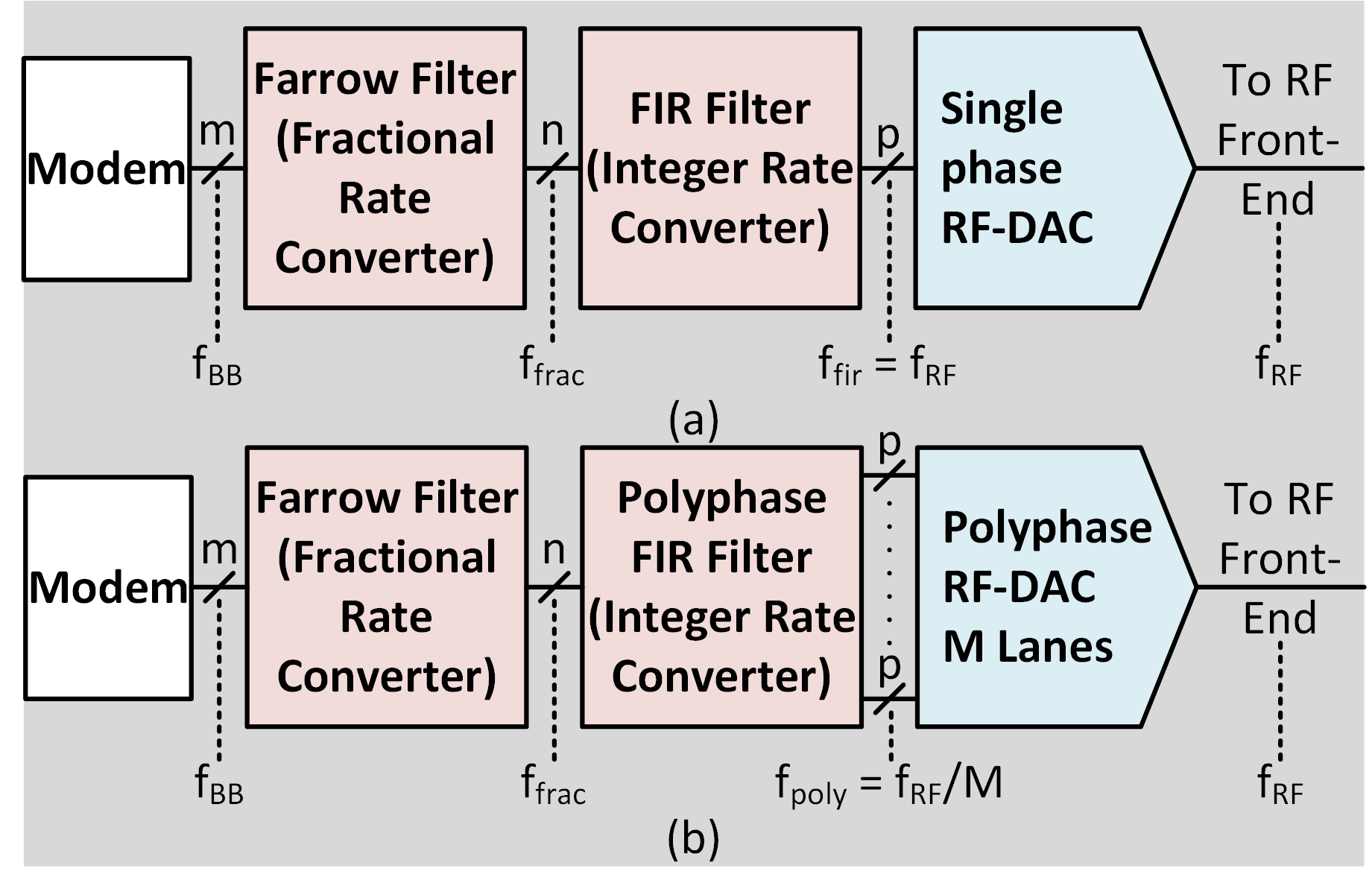}
  \vspace{-5mm}
  \caption{RF-DAC interpolation architectures. (a)~Conventional direct-digital FIR: requires logic running at $f_{RF}$. (b)~Polyphase hybrid analog-digital FIR: $M$ lanes, each driving an RF-DAC slice, reduce the maximum digital rate to $f_{RF}/M$; this paper implements $M=4$ and $M=8$.}
  \label{fig:top_level}
  \vspace{-6mm}
\end{figure}

However, in both commercial and military networks, higher frequency spectrum, such as that used in both \ac{FR2} and \ac{FR3} bands is being proposed to offset the severe congestion that exists in \ac{FR1} bands \cite{shakya2024_fr3}. A key challenge to scaling hardware systems to higher frequency spectrum is that the \ac{BB} signals typically use wider bandwidths and in the case of \acp{RF-DAC} the \ac{BB} signals must be interpolated to higher sample rates to help suppress spectral images that arise from the process of sampling.

In conventional RF wireless systems (e.g., digital-\ac{BB} + \ac{DAC} + \ac{RF} front-end), the sampling images can be suppressed using active, high-order filters that serve as the reconstruction filter of the DAC. These are impractical when using an \ac{RF-DAC}, where the reconstruction filter is typically a low-order passive \ac{RF} filter acting as an impedance matching network. Instead, the sample rate must be increased as much as possible, and ideally as high as the output frequency, which enables easier suppression of the images. This is similar to the concept of oversampling in \ac{BB} \acp{DAC}.

Advanced CMOS process nodes are capable of supporting high sampling rates~\cite{bhide2015interleaved,olieman2015_interleaved, caragiulo2021ti}. However, these are still traditional baseband architectures that require separate up-conversion chains and hence do not offer the flexibility of the \ac{RF-DAC}. Furthermore, the designs feature highly customized digital architectures to meet stringent timing requirements. Time-interleaved \mbox{$\Delta\Sigma$} {RF-DAC} architectures~\cite{mccue2016tidsm} and direct-RF \acp{TX} with embedded FIR-like impulse responses~\cite{su1993_oversampling, barkin2004_oversampling, taleie2006_fir, bhat2017_fir, su2020timeapprox, su2021_taf} have explored related techniques, but remain constrained in filter reconfigurability and achieve limited tap depth. 

\begin{figure*}[b]
  \centering
  \includegraphics{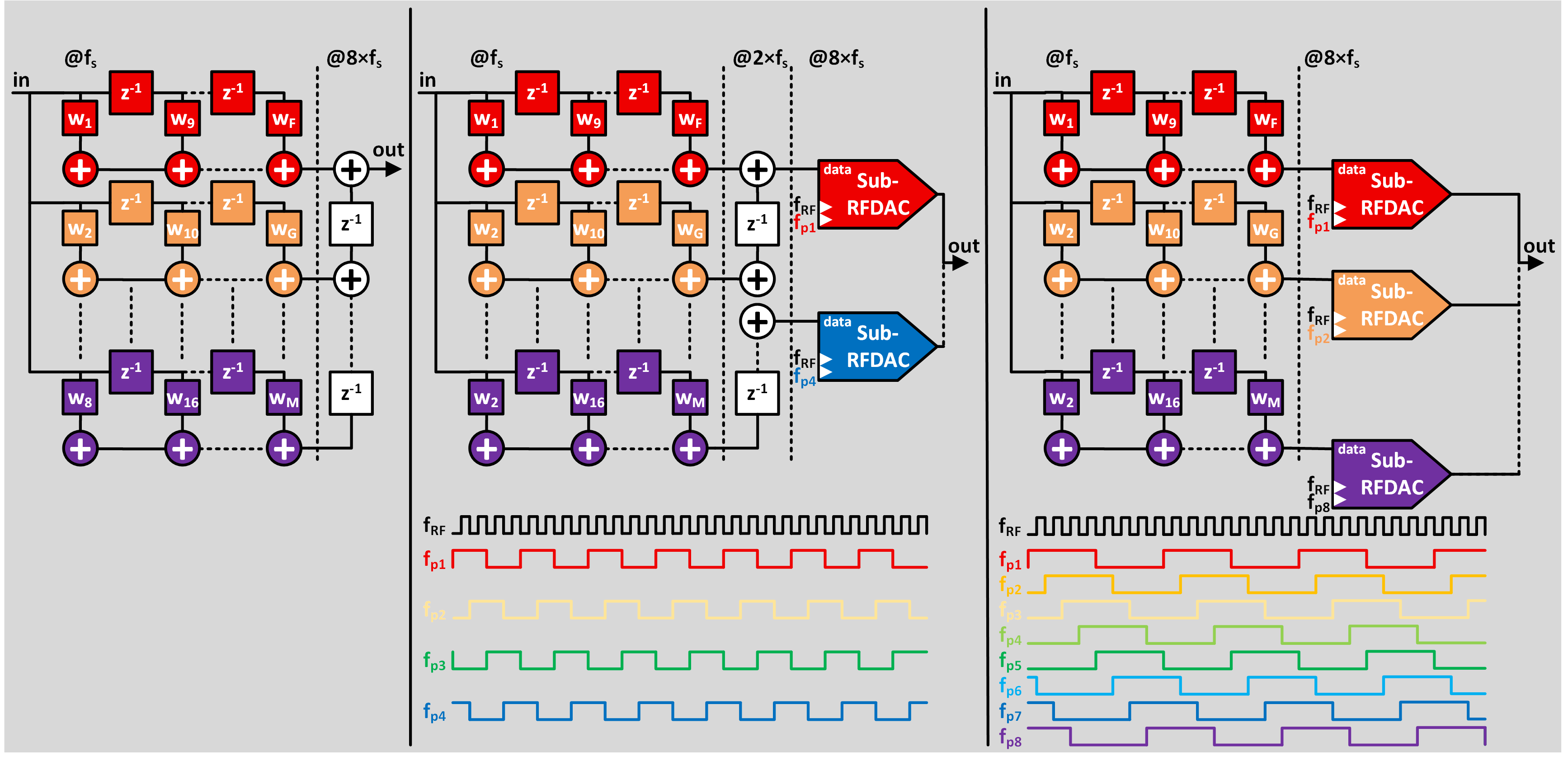}
  \vspace{-3mm}
  \caption{Detailed polyphase interpolation filter block diagrams. \textbf{(left)}~Conventional polyphase FIR integer rate resampling, with recombination in the digital domain. Output delay and sum operations run at $f_{RF}$. \textbf{(middle)}~4-phase polyphase decomposition: eight sub-filters running at $f_\mathrm{RF}/8$, before parallel branches are recombined at $f_\mathrm{RF}/4$, each driving a dedicated RF-DAC slice. \textbf{(right)}~8-phase polyphase decomposition: eight sub-filters running at $f_\mathrm{RF}/8$, each driving a dedicated RF-DAC slice.}
  \label{fig:polyphase_detail}
  \vspace{-5mm}
\end{figure*}

The challenge in purely analog interpolation methods is that they provide only a limited number of filter taps. In a \ac{TI} converter array~\cite{black1980_ti}, the number of taps equals the number of interleaved phases. In a \ac{TAF}, the limitation is on the available shaping of the clock pulse~\cite{su2020timeapprox,su2021_taf}. Additionally, the shaping on the clock pulse can have other adverse effects such as spur generation. In analog FIR filters, the limitation is on the number of individual analog processing cores~\cite{taleie2006_fir,bhat2017_fir}. This limits the \ac{OOB} rejection and adaptability to new waveforms. Hybrid approaches, as shown in \Cref{fig:top_level} that merge analog and digital filtering capabilities, are an alternative approach. A single-phase \ac{RF-DAC} with a digital interpolation filter would rely on the filter operating at output speeds up to the \ac{RF} carrier frequency for maximal image suppression. This is practical if operation is limited to \ac{FR1} frequencies (below $7$\,GHz), but becomes impractical as frequencies scale into \ac{FR3} ($7$--$24$\,GHz) and \ac{FR2} (above $24$\,GHz), where even full-custom digital FIR filters have only approached $10$\,GHz~\cite{muller2012_7bit}. 

This paper shows that the recombination of a polyphase interpolator can be split between digital commutators and the RF-DAC itself, giving either partial digital recombination followed by analog recombination, or fully analog recombination, and implements one variant of each (\Cref{fig:polyphase_detail}, middle and right). A Farrow fractional-rate filter in software/FPGA decouples the fixed baseband data rate from the RF clock, and the multi-GHz integer-rate interpolation is decomposed into polyphase form on a dedicated mixed-signal ASIC macro. The 4-phase variant needs only four clock phases, easily derived from the \ac{RF} local oscillator, but its digital logic must close timing at $f_\mathrm{RF}/4$; the 8-phase variant needs a more complex multiphase clock network but halves the digital rate to $f_\mathrm{RF}/8$.

\Cref{sec:arch} describes the generalized recombination and the system partitioning. In \Cref{sec:analysis} the recombination model, along with its mismatch, clock and in-band noise and distortion budgets is analyzed. Next, \Cref{sec:choosingM} provides guidance on the optimal choice of the number of clock phases, $M$. Implementation and post-layout results are presented in \Cref{sec:implementation,sec:results}. Finally, \Cref{sec:conclusion} concludes the paper.
\section{System Architecture and Partitioning}
\label{sec:arch}

The block diagram of \Cref{fig:top_level}(b), and the detailed signal-flow graphs of \Cref{fig:polyphase_detail}, illustrate the proposed hybrid architectures for implementing a high-rate interpolation filter for an RF-DAC output. To achieve frequency agility while maintaining physical implementation viability, the architecture employs a distinct partitioning. The system relies on an initial Farrow filter structure for efficient fractional-rate resampling of the baseband signal to a rate that is an integer fraction of the desired \ac{RF} output frequency. Farrow filters are well-suited for this role due to their ability to perform arbitrary-rate interpolation with a fixed set of filter coefficients~\cite{farrow1988continuously,hentschel2000_sample}. Given its mathematical complexity and requirements for runtime parameter adjustment, this fractional block is designated for flexible off-chip deployment (e.g., on an FPGA front-end).

The output of this off-chip Farrow filter then feeds the custom mixed-signal ASIC macro, which houses a polyphase FIR bank that generates multiple parallel hardware output streams, each operating at a manageable fraction of the target output frequency. As shown in \Cref{fig:polyphase_detail} (left), a pure-digital polyphase decomposition requires a digital commutator running at $f_\mathrm{RF}$ to reconstruct the output sequence. The proposed hybrid architectures move all or part of this recombination into the RF-DAC, where it is performed passively. This can be done using either current- or voltage-mode RF-DACs, by summing currents \cite{hashemi2017_intrinsically} or by controlling charge division on capacitor arrays \cite{yuan2017_multiphase}, with the recombination controlled by interleaved clock phases.

\subsection{Generalized Digital/Analog Polyphase Recombination}
\label{sec:general}

A length-$N$ FIR filter $H(z)$ with coefficients $h[n]$ is decomposed into $P$ polyphase sub-filters~\cite{crochiere1983multirate, vaidyanathan1993multirate, harris2004multirate},
\begin{equation}
  H(z) = \sum_{k=0}^{P-1} z^{-k} E_k(z^P),
  \label{eq:polyphase_P1}
\end{equation}
where each sub-filter, with $h[n]=0$ for $n \ge N$, is defined as:
\begin{equation}
  E_k(z) = \sum_{n=0}^{\lceil N/P \rceil - 1} h[nP + k]\, z^{-n}.
  \label{eq:polyphase_P2}
\end{equation}
Each sub-filter runs at $f_\mathrm{RF}/P$. The recombination in \eqref{eq:polyphase_P1} is what must run at $f_\mathrm{RF}$. Set $D = P/M$, and index the sub-filters as $k = \ell + Md$ with $\ell = 0,\ldots,M-1$ and $d = 0,\ldots,D-1$. The same filter then factors into two nested recombinations,
\begin{equation}
  H(z) = \sum_{\ell=0}^{M-1} z^{-\ell}\, F_\ell(z^M),
  \label{eq:two_level1}
\end{equation}
where $F_\ell(z)$, the $\ell$-th $M$-phase component of $H(z)$, is assembled from $D$ of the sub-filters:
\begin{equation}
  F_\ell(z) = \sum_{d=0}^{D-1} z^{-d}\, E_{\ell+Md}(z^D).
  \label{eq:two_level2}
\end{equation}
The sum in \eqref{eq:two_level2} interleaves $D$ sub-filters into one stream at $f_\mathrm{RF}/M$: a $D$-way digital commutator. The sum in \eqref{eq:two_level1} interleaves the $M$ streams $y_\ell[r]$ produced by the $F_\ell$ into the full-rate output, and it is this stage that the RF-DAC performs, with one slice per stream and $M$ clock phases spaced $360^\circ/M$ apart at $f_\mathrm{RF}/M$~\cite{yuan2017_multiphase,khoini1997_time}. The pair $(P, M)$ thus spans a design space whose regimes are shown in the three columns of \Cref{fig:polyphase_detail}:
\begin{itemize}
  \item $M = 1$ ($D = P$): \emph{fully digital} recombination. A single
        commutator runs at $f_\mathrm{RF}$ and drives one DAC
        (\Cref{fig:polyphase_detail}, left).
  \item $1 < M < P$: \emph{partial digital} recombination. $M$ commutators
        run at $f_\mathrm{RF}/M$, and their outputs are recombined in the
        analog domain by $M$ RF-DAC slices (\Cref{fig:polyphase_detail},
        middle, $P=8$, $M=4$).
  \item $M = P$ ($D = 1$): \emph{fully analog} recombination. Every
        sub-filter drives its own slice with no digital commutators
        (\Cref{fig:polyphase_detail}, right, $P = M = 8$).
\end{itemize}

\begin{table}[t]
\caption{Rates and Resources of the Generalized Recombination, $P=8$, $f_\mathrm{RF}=12.8$\,GHz}
\label{tab:general}
\centering
\resizebox{\columnwidth}{!}{
\begin{tabular}{lcccc}
\hline
 & $M=1$ & $M=2$ & $M=4$ & $M=8$ \\
 & (digital) & & (this work) & (this work) \\ \hline
Sub-filter rate $f_\mathrm{RF}/P$ & $1.6$\,GHz & $1.6$\,GHz & $1.6$\,GHz & $1.6$\,GHz \\
Digital commutators $\times$ ways & $1\times 8$ & $2\times 4$ & $4\times 2$ & none \\
Highest digital rate $f_\mathrm{RF}/M$ & $12.8$\,GHz & $6.4$\,GHz & $3.2$\,GHz & $1.6$\,GHz \\
RF-DAC slices & $1$ & $2$ & $4$ & $8$ \\
Clock phases (spacing) & --- & $2$ ($180^\circ$) & $4$ ($90^\circ$) & $8$ ($45^\circ$) \\
Analog hold $T_b = M T_\mathrm{RF}$ & $78.1$\,ps & $156$\,ps & $312.5$\,ps & $625$\,ps \\ \hline
\end{tabular}
}
\vspace{-4mm}
\end{table}

\Cref{tab:general} lists the resulting rates. It is noteworthy that the digital commutators are synchronous logic whose correct timing is guaranteed by static timing analysis at sign-off, so they contribute no amplitude or timing error of their own, and every analog
non-ideality (e.g., the staggered hold, the slice mismatch and the clocks of \Cref{sec:analysis}) depends on $M$ alone. Additionally, the fastest digital clock required is $f_\mathrm{RF}/M$ for any $P$, so it is $M$, and not $P$, that sets the required digital timing closure; $P \ge M$ only sets how parallel the FIR arithmetic is \cite{tsao2012_area}. The optimal choice of $M$ is analyzed and quantified in \Cref{sec:choosingM}.

\subsection{Polyphase FIR Architecture}
\label{sec:polyphase}

To demonstrate the effectiveness of the proposed techniques, we designed an interpolation filter targeting 802.11be-like $320$\,MHz signals at \ac{FR3} frequencies. Though no monolithic standard currently exists, this particular combination of wide signal bandwidth ($320$\,MHz) and ultra-high output frequency ($12.8$\,GHz nominal, up to $16$\,GSps signed off) serves to demonstrate the capabilities of the proposed technique for high-dynamic-range, wideband signals.

\subsubsection{Filter Specification}
The integer-rate interpolation from $f_\mathrm{RF}/32$ to $f_\mathrm{RF}$ is split into a $\times 4$ pre-interpolator, from $f_\mathrm{RF}/32$ to $f_\mathrm{RF}/8$, followed by a $\times 8$ stage. The $\times 4$ pre-interpolator is designed to a $160$\,MHz passband, a $240$\,MHz stopband, and $> 80$\,dB stopband attenuation (before coefficient quantization). The $\times 8$ stage keeps the same $160$\,MHz passband and rejects the images of the $f_\mathrm{RF}/8$ stream, centered at multiples of $1.6$\,GHz.

To achieve these characteristics, the $\times 8$ stage cascades a $\times 2$ FIR filter, which compensates the passband droop, with the response of a 6-stage $\times 4$ \ac{CIC} filter \cite{hogenauer1981_cic}. The $\mathrm{sinc}^6$ response of the CIC places its nulls at the image centers; its droop is removed by the compensation filter, which also enforces the stopband attenuation \cite{kwentus1997_droop}. No CIC structure is built: the cascade is implemented as an equivalent direct-form FIR for polyphase decomposition.

\subsubsection{Polyphase Decomposition and Coefficient Quantization}

The coefficients of the equivalent $\times 8$ cascade were derived using the impulse response of the full structure, quantized to 16 bits for hardware execution, and decomposed according to \eqref{eq:polyphase_P1} and \eqref{eq:polyphase_P2} with $P = 8$, giving $165$ taps. The $\times 4$ pre-interpolator is a $95$-tap Nyquist(4) filter, also with 16-bit coefficients, implemented as four polyphase branches. After quantization, the pre-interpolator attenuates its stopband by $66.8$\,dB, and the $\times 8$ cascade attenuates the images of the $f_\mathrm{RF}/8$ stream by more than $90$\,dB, both well above the $55$\,dBc \ac{SFDR} target of \Cref{sec:analysis}.

To enable a rigorous comparative analysis between the 4- and 8-phase physical macro variants, we utilize a single set of master filter coefficients and apply the same polyphase decomposition framework to both. The 4-phase architecture combines interleaved pairs of the 8-phase branches in the digital domain, i.e., \eqref{eq:two_level1} and \eqref{eq:two_level2} with $D=2$, while the 8-phase architecture directly implements all eight physical hardware branches. This approach ensures both variants maintain an identical underlying filter response, isolating the post-layout evaluation to differences in digital logic power, clock routing, and layout congestion.

For the $M=8$ hardware instantiation, the input bitstream data is fed in parallel to all eight sub-filters on the macro, each operating at $1.6$\,GHz and outputting one sample per clock cycle. These eight outputs map to successive time offsets $k T_\text{RF}$ ($k = 0, 1, \ldots, 7$, where $T_\text{RF} = 1/f_\text{RF}$). These parallel streams directly drive the input ports of the eight RF-DAC slices, sequencing their updates across the $45^\circ$-spaced clock phases to passively reconstruct the $12.8$\,GHz analog output waveform.

\subsection{Architecture Trade-offs}

 A single FIR at $f_\mathrm{RF} = 12.8$ GHz (\Cref{fig:top_level}(a)) exceeds what even full-custom digital FIR filters have achieved; these have only approached $10$ GHz, by relying on power-of-two coefficients, bit-level pipelining and supply overdrive~\cite{muller2012_7bit}. A fully digital polyphase decomposition ($M=1$) still needs an output commutator at $f_\mathrm{RF}$. Both implemented variants avoid these constraints. The 4-phase variant ($P=8$, $M=4$) needs four $90^\circ$-spaced phases at $f_\mathrm{RF}/4$, which a divide-by-4 counter clocked by $f_\mathrm{RF}$ provides directly, but its four two-way commutators must close timing at $3.2$\,GHz. The 8-phase variant ($P=M=8$) needs eight $45^\circ$-spaced phases, a more complex clock network, in exchange for removing the commutators, so its fastest digital clock is the $1.6$\,GHz sub-filter rate.
\section{Recombination Model and Mismatch Effects}
\label{sec:analysis}

\subsection{Staggered Zero-Order Hold}

This section analyzes the analog stage of \eqref{eq:two_level1}, the recombination of the $M$ streams on the RF-DAC. As argued in \Cref{sec:general}, the digital commutators contribute no error of their own, so every result below depends on $M$ alone and holds for any $D = P/M$, i.e., for partial and fully analog recombination alike.

The recombination performed on the slice array is not a one-hot commutator. On each slice update-clock edge all $M$ stream words are held, then loaded into the array sequentially, one per successive $f_\mathrm{RF}$ cycle. Once loaded, a slice continues to switch at $f_\mathrm{RF}$ with that data until its next update, so every slice drives the array for a full slice update period $T_b = M T_\mathrm{RF}$. If the slice weights are equal, the slices differ only in \emph{when} their data changes, and the array output is a staggered zero-order hold,
\begin{equation}
  y(t)=\sum_{k=0}^{M-1}\sum_{r} y_k[r]\,
       p_{T_b}\left(t-rT_b-kT_\mathrm{RF}\right),
  \label{eq:staggered_zoh}
\end{equation}
where $y(t)$ is the baseband-equivalent output, $p_{T_b}(\cdot)$ is a $T_b$-wide unit pulse, and $y_k[r]$ is the stream that drives slice~$k$, i.e., the output of $F_\ell$ in \eqref{eq:two_level2} with $\ell = k$. Since $y_k[r] = v[rM+k]$, where $v[n]$ is the output of the full-rate filter $H(z)$, substituting $n = rM+k$ in \eqref{eq:staggered_zoh} gives
\begin{equation}
  y(t)=\sum_{n} v[n]\,p_{T_b}\left(t-nT_\mathrm{RF}\right).
  \label{eq:wide_zoh}
\end{equation}

\subsubsection{Hold Response}
By \eqref{eq:wide_zoh}, the array is equivalent to a single DAC updated at $f_\mathrm{RF}$ whose zero-order hold lasts $T_b = M T_\mathrm{RF}$ instead of $T_\mathrm{RF}$ \cite{balasubramanian2011systematic,duncan2017_dac}. Its frequency response is the usual sinc of a zero-order hold, $P_{T_b}(f) = T_b\,\mathrm{sinc}(fT_b)$, but with nulls at multiples of $f_b = f_\mathrm{RF}/M$ rather than of $f_\mathrm{RF}$; at RF they lie at $\pm m f_b$ from the carrier. The $M$-times narrower sinc also increases the passband droop, but only to $0.04$\,dB ($M=4$) and $0.14$\,dB ($M=8$) at the $160$\,MHz band edge, which the FIR can pre-compensate if required. A sample-level simulation of the slice array agrees with \eqref{eq:wide_zoh} to within $2\times10^{-15}$, i.e., double-precision rounding, for both variants.

\subsubsection{Effective Resolution}
\label{sec:resolution}
Equation~\eqref{eq:wide_zoh} also bounds the achievable resolution. Slices carrying successive time samples add full-scale output and update rate but never subdivide the unit element, so $M$ slices of $B_s$ bits do \emph{not} yield $B_s+\log_2 M$ bits. Each slice quantizes independently, so the in-band \ac{SNDR} improves over a single slice by only $10\log_{10} M$, i.e., $\tfrac{1}{2}\log_2 M$ bits ($+1.0$\,bit for $M=4$, $+1.5$\,bit for $M=8$), which simulation confirms. The array therefore falls $\tfrac{1}{2}\log_2 M$ bits short of a $(B_s+\log_2 M)$-bit DAC, even with perfectly matched slices. \Cref{sec:spectrum} treats unequal slice weights and timing skew, whose spurs at $f_0 + m f_b$ fall on the skirts of the sinc nulls.

\subsection{Impact of Timing and Gain Mismatch on Spurious Outputs}
\label{sec:spectrum}

With a per-slice gain error $g_k = 1+\epsilon_k$ and timing error/skew
$\tau_k$, the output spectrum is
\begin{equation}
  Y(f)=P_{T_b}(f)\sum_{k=0}^{M-1} g_k A_k(f)\,
       e^{-j2\pi f\left(kT_\mathrm{RF}+\tau_k\right)},
  \label{eq:spectrum}
\end{equation}
where $A_k(f)$ is the branch spectrum, that is periodic with $f_b = 1/T_b$, and $P_{T_b}(f)=T_b\,\mathrm{sinc}(fT_b)$. Because every slice holds for exactly $T_b$, $P_{T_b}(f)$ factors out of the sum. Hence, mismatch never moves the location of the null, it only reduces the cancellation it achieves across the null's skirts.

For a single tone at $f_0$ each $A_k(f)$ is a comb at $f_m = f_0 + m f_b$. Within $\pm f_\mathrm{RF}/2$ of the carrier, the output has only the carrier ($m=0$) and the polyphase spurs $m = -M/2,\ldots,M/2$, $m \neq 0$. With $x \triangleq f_0 T_b$, the hold weights the spurs relative to the carrier by the exact ratio:
\begin{equation}
  \frac{P_{T_b}(f_m)}{P_{T_b}(f_0)} = (-1)^m\,\frac{x}{m+x}.
  \label{eq:hold_ratio}
\end{equation}
To a first order in $\epsilon_k$ and $\tau_k$, the spur-to-carrier amplitude is then:
\begin{equation}
  \frac{Y(f_m)}{Y(f_0)} \approx \frac{(-1)^m}{M}
  \left[\frac{x}{m+x}\,\mathcal{E}_q - j2\pi f_0\,\mathcal{T}_q\right],
  \label{eq:spur_dft}
\end{equation}
where $q = m \bmod M$, $W_M = e^{-j2\pi/M}$, and $\mathcal{E}_q=\sum_k \epsilon_k W_M^{qk}$ and $\mathcal{T}_q=\sum_k \tau_k W_M^{qk}$ are the $M$-point \acp{DFT} of the slice gain and timing errors \cite{jenq1988_dacspectra,kurosawa2001_tidac}. A skew $\tau_k$ gives slice $k$ a phase error of $2\pi f_m \tau_k$ at the spur frequency $f_m$, and the hold attenuates that spur by $x/(m+x)$ from \eqref{eq:hold_ratio}. Since $f_m\,x/(m+x) = f_0$, the product is $2\pi f_0 \tau_k$ for every $m$, so the timing sensitivity depends on the signal frequency $f_0$ and not on $f_\mathrm{RF}$ \cite{elchammas2009_phase}. Skew of the $f_\mathrm{RF}$ clock itself behaves differently, because it also rotates the carrier of the slice; \Cref{sec:clocks} shows that its per-spur weight is $M/(1-x)$ times larger, i.e., $12.5$\,dB at $M=4$ and $19.0$\,dB at $M=8$, which tightens the median skew budget by $10.6$ and $14.9$\,dB (\Cref{tab:clock}). The gain error is attenuated by $x/(m+x)$ alone, least at $m=-1$, where $|m+x| = 1-x$ is smallest.

\label{sec:spur_stats}%
To turn \eqref{eq:spur_dft} into a budget, let the slice errors be independent, zero-mean Gaussian random variables. The gain (or weighting) error is $\epsilon_k \sim \mathcal{N}(0,\sigma_g^2)$, while the timing error is $\tau_k \sim \mathcal{N}(0,\sigma_\tau^2)$, where $\sigma_g$ and $\sigma_\tau$ are the respective standard deviations of the errors. For a real white Gaussian sequence, the \ac{DFT} bins $q = 1,\ldots,M/2$ are mutually independent, and bin $M-q$ is the complex conjugate of bin $q$. The spurs at $m = \pm q$ therefore have the same random amplitude, and we refer to them as group $q$. For $q < M/2$ the bin is a circular complex Gaussian of variance $M\sigma^2$, so the spur power is exponentially distributed. The bin $q = M/2$ is real, so its spur power is a scaled $\chi^2_1$ variable. Taking the larger spur of each group, the mean spur power of group $q$ is
\begin{align}
  \mu_q^{(\tau)} &= \frac{\left(2\pi f_0\sigma_\tau\right)^{2}}{M},
  \label{eq:timing_law}\\
  \mu_q^{(g)} &= \frac{\sigma_g^{2}}{M}\cdot\frac{x^{2}}{(q-x)^{2}},
  \label{eq:gain_law}
\end{align}
and the two mechanisms add in power. Timing skew contributes equally to every group, whereas gain error falls mainly on $q=1$. The \ac{SFDR} is set by the worst spur. Because the groups are independent, its cumulative distribution is the product of the group distributions,
\begin{equation}
  F(t) = \prod_{q=1}^{M/2-1}\left(1-e^{-t/\mu_q}\right)\cdot
         \mathrm{erf}\sqrt{\frac{t}{2\mu_{M/2}}}.
  \label{eq:cdf_max}
\end{equation}
For a single mechanism, every $\mu_q$ is proportional to $\sigma^2$, where $\sigma$ is either $\sigma_\tau$ or $\sigma_g$. Let $\eta$ denote the yield, i.e., the fraction of fabricated chips whose worst spur lies below a level $t_\eta$, so that $F(t_\eta)=\eta$. This level can then be written as $t_\eta=\kappa_\eta\sigma^2$, and a yield $\eta$ at an \ac{SFDR} target of $S$\,dBc is achieved if \cite{leger2004_impact}:
\begin{equation}
  \sigma \le \sigma_\mathrm{max}(S,\eta) = \sqrt{10^{-S/10}/\kappa_\eta}.
  \label{eq:yield_law}
\end{equation}
The yield cost in \ac{SFDR} is the gap between $t_\eta$ and the largest group mean $\bar{\mu} = \max_q \mu_q$. For gain error, $q=1$ dominates and the worst spur behaves like a single exponential variable: it lies $1.1$--$1.4$\,dB below $\bar{\mu}$ at the median and $4.8$\,dB above it at $95\%$ yield. For timing skew, all $M/2$ groups have the same mean, so the worst of them grows with $M$: $0.7$ and $2.7$\,dB above $\bar{\mu}$ at the median for $M=4$ and $8$, and $6.4$ and $6.9$\,dB at $95\%$.

A given SFDR target follows from the slice resolution of the subsequent RF-DACs. In our example each slice is a sign-magnitude DAC with an $8$-bit magnitude ($255$ unary units plus sign, i.e., $511$ levels), whose quantization-limited \ac{SNR} for a full-scale sine is $55.9$\,dB. Recombining $M$ slices adds $10\log_{10} M$, as shown in \Cref{sec:resolution}, i.e., $61.9$\,dB at $M=4$ and $64.9$\,dB at $M=8$, over the $f_b$ Nyquist band for a full-scale sine. Hence, a reasonable limit for the \ac{SFDR} target is set by the quantization-limited SNR of a single slice. For example, mismatch spurs at or below $55$\,dBc remain below the quantization noise of any one slice. For a $55$\,dBc \ac{SFDR} with the tone at the band edge, $f_0 = 160$\,MHz, \eqref{eq:yield_law} gives a median budget of $\sigma_\tau \le 3.26$\,ps and $\sigma_g \le 7.7\%$ at $M=4$, and $\sigma_\tau \le 3.66$\,ps and $\sigma_g \le 5.3\%$ at $M=8$. At $95\%$ yield the budgets tighten to $1.70$ and $2.27$\,ps, and $3.9\%$ and $2.6\%$. For a different target $S$, e.g., with higher-resolution slices, the error budgets scale by $10^{-(S-55)/20}$, i.e., they halve for every $6$\,dB of additional \ac{SFDR}. To verify the closed-form estimates, a $2\times10^4$-trial Monte Carlo of the exact, non-linearized \eqref{eq:spectrum} was run, and it agrees with them to within $0.1$\,dB for every \ac{SFDR} above $45$\,dBc; beyond $\sigma_\tau \approx 7$\,ps the first-order expansion becomes conservative, by at most $1.25$\,dB at $20$\,ps. The yield law also explains why doubling $M$ relaxes the median timing budget by only $12\%$ rather than the $\sqrt{2}$ that the single-spur mean \eqref{eq:timing_law} suggests: the $1/M$ reduction of each spur is largely offset by the larger number of independent spurs competing for the worst case. The gain budget tightens with $M$ because $x = f_0 T_b$ doubles.

In the case of a modulated stimulus of bandwidth $B$, each spur is spread over the band, at baseband offsets $\nu$ between $-B/2$ and $B/2$. Averaging $\nu^2$ over a flat band gives $\langle\nu^2\rangle = (B/2)^2/3$, so to leading order $f_0$ in \eqref{eq:timing_law} and \eqref{eq:gain_law} is replaced by $B/2$, and the spur power drops by a further factor of $3$. For gain this is $25$--$31$\,dB below the conventional interleaving result $\sigma_g^2/M$. A Monte Carlo simulation with a $320$\,MHz modulated stimulus, evaluated on the worst spur band, gives budgets of $5.7$\,ps ($M=4$) and $6.6$\,ps ($M=8$) for timing skew and $14.5\%$ and $10.0\%$ for gain error. These are all looser than the single-tone budgets above, hence the band-edge tone sets the requirement.

To summarize, inter-slice amplitude matching is not the critical constraint; the $f_\mathrm{RF}$ clock arrival at the slices is (\Cref{sec:clocks}), while intra-slice unit matching sets the in-band SNDR (\Cref{sec:unitcap,sec:sndr}).

To this point, the skew $\tau_k$ considered is only at the instant when a slice's data changes, and \Cref{sec:clocks} shows that it is not the dominant timing error.

\subsection{Clock Consideration on the Timing Budget}
\label{sec:clocks}

Each slice is timed by two clocks: its phase $\phi_k$ of the slice update clock, at $f_b$ and delayed by $kT_\mathrm{RF}$, which delivers its new input code, and the $f_\mathrm{RF}$ clock that switches its current sources or capacitors. In the proposed implementation the baseband word is retimed by the $f_\mathrm{RF}$ clock at the slice, so the update instant follows the local $f_\mathrm{RF}$ edge and $\phi_k$ only has to land inside the retiming window.
$\tau^{c}_k$ is the arrival skew of the $f_\mathrm{RF}$ clock at slice~$k$ \cite{razavi2013_interleaving}. The slice output $b_k(t-\tau^{c}_k)\,c(t-\tau^{c}_k)$, with $c(t)$ the carrier, has the baseband equivalent $b_k(t-\tau^{c}_k)\,e^{-j2\pi f_\mathrm{RF}\tau^{c}_k}$: besides delaying the data, the skew rotates the slice's carrier by a constant phase across the hold duration. A carrier rotation is a complex slice gain, so the skirt weights it like $\epsilon_k$, and \eqref{eq:spur_dft} becomes
\begin{equation}
  \frac{Y(f_m)}{Y(f_0)} \approx \frac{(-1)^m}{M}
  \left[\frac{x}{m+x}\left(\mathcal{E}_q - j\Theta_q\right)
        - j2\pi f_0\,\mathcal{T}_q\right],
  \label{eq:spur_carrier}
\end{equation}
with $\Theta_q$ the \ac{DFT} of $\theta_k = 2\pi f_\mathrm{RF}\tau^{c}_k$. A carrier skew therefore acts as a quadrature gain error of $\sigma_\theta = 2\pi f_\mathrm{RF}\sigma_c$, and its budget is the gain budget divided by $2\pi f_\mathrm{RF}$. Its weight $2\pi f_0\,M/(m+x)$ exceeds that of the data-update term by $M/(1-x)$, i.e., $12.5$\,dB at $M=4$ and $19.0$\,dB at $M=8$. With retiming, $\mathcal{T}_q = \Theta_q/(2\pi f_\mathrm{RF})$ and the two terms combine into the weight $2\pi f_0\,(M+m+x)/(m+x)$ of a whole-slice time shift, worst at $m=+1$. The phase clocks $\phi_k$, in contrast, cause no spur when their skew stays inside the retiming window $w \le T_\mathrm{RF}/2$, and all $M$ slices clear it with probability $\eta$ for
\begin{equation}
  \sigma_\phi \le \frac{w}{\Phi^{-1}\!\left(\tfrac{1}{2}\left(1+\eta^{1/M}\right)\right)}.
  \label{eq:slip_law}
\end{equation}
Beyond this timing window, a slice will latch the wrong $f_\mathrm{RF}$ cycle and the worst image rises to $-40.0$\,dBc at $M=4$ and $-45.4$\,dBc at $M=8$.

\begin{table}[tbh]
\caption{RMS Skew for $55$\,dBc \ac{SFDR}, Single Tone at $f_0 = 160$\,MHz, Median / $95\%$ Yield (ps)}
\label{tab:clock}
\centering
\begin{tabular}{lcc}
\hline
Skew source & $M=4$ & $M=8$ \\ \hline
Data update, not retimed & $3.26$ / $1.70$ & $3.66$ / $2.27$ \\
$f_\mathrm{RF}$ carrier only & $0.96$ / $0.48$ & $0.66$ / $0.33$ \\
$f_\mathrm{RF}$ arrival, retimed$^\ast$ & $0.79$ / $0.42$ & $0.67$ / $0.35$ \\
Phase clock $\phi_k$, retimed$^\ast$ & $27.7$ / $15.7$ & $22.5$ / $14.3$ \\ \hline
\multicolumn{3}{l}{\scriptsize $^\ast$Assumed in the proposed design (retiming at each slice);}\\
\multicolumn{3}{l}{\scriptsize $\phi_k$ from \eqref{eq:slip_law} with $w = T_\mathrm{RF}/2$.}
\end{tabular}
\vspace{-4mm}
\end{table}

Shown in \Cref{tab:clock} are the analytically obtained timing skew error budgets to achieve the $55$\,dBc \ac{SFDR} target. Each agrees to within $1\%$ with a $2\times10^4$-trial Monte Carlo of \eqref{eq:spectrum} in which each slice gain $g_k$ also includes the carrier phase $e^{-j2\pi f_\mathrm{RF}\tau^{c}_k}$. What is observed is that the slice update clocks are not the critical elements. Because each slice retimes its data on the local $f_\mathrm{RF}$ edge, in the median case, the $M=4$ and $M=8$ slice update clocks can be skewed by up to $28$ and $23$\,ps rms before a slice latches the wrong $f_\mathrm{RF}$ cycle. The critical clock is the $f_\mathrm{RF}$ clock itself, because its skew also rotates the carrier of each slice. Its arrival at the slices must match to $0.79$\,ps ($M=4$) and $0.67$\,ps ($M=8$) rms at the median, and to $\approx0.4$\,ps for $95\%$ yield. Only differential skew between the slices is consequential; common skew to all of the slices does not result in a spur. Since the carrier rotation is a fixed phase per slice, it could be measured and removed by a digital phase rotation of each stream, as in the background calibration of time-interleaved ADCs~\cite{elbornsson2004_blind}. In that case, only the data-update error would remain, with the few-picosecond budgets of \eqref{eq:timing_law}.

In degrees of the slice update clock, a skew $\tau$ is $\delta = 360^\circ f_b\,\tau$. The phase-clock budgets of \Cref{tab:clock} are therefore $32^\circ$ ($M=4$) and $13^\circ$ ($M=8$) rms at the median, and $18^\circ$ and $8.3^\circ$ at $95\%$ yield; without retiming they would be $3.8^\circ$ and $2.1^\circ$. Since $f_b = f_\mathrm{RF}/M$, the fixed slip window shrinks in degrees as $M$ grows, and the $45^\circ$ generator must be $1.8$ to $2.5$ times more accurate than the $90^\circ$ one.

\subsection{Unit-Capacitor Mismatch}
\label{sec:unitcap}

Section~\ref{sec:spectrum} treats each slice as a single weight with gain error $\epsilon_k$. Inside a slice, however, the weight is built from unit capacitors, and their mismatch produces a code-dependent error. In the RF-DAC assumed for this design, each slice has a sign-magnitude I- and Q-array \cite{mehrpoo2018_wideband}, each with $N_u$ thermometer-coded unit capacitors; the implemented macros drive $N_u = 255$ units per array through their binary-to-thermometer decoders. Unit $i$ of array $c \in \{\mathrm{I},\mathrm{Q}\}$ in slice $k$ has a fixed relative error $\sigma_u e_{k,c,i}$, where the $e_{k,c,i} \sim \mathcal{N}(0,1)$ are independent and $\sigma_u = \sigma_{\Delta C/C}$ is the rms unit-capacitor mismatch \cite{shyu1984_error}.

Unit $i$ is switched on whenever the code magnitude is at least $i$, and it then contributes $\pm 1/N_u$ according to the sign of the code. Its error therefore adds $\sigma_u e_{k,c,i}$ times this on/off pattern to the output, and the output spectrum at frequency bin $j$ is
\begin{equation}
  Y_j = Y_j^{(0)} + \sigma_u \sum_{k,c,i} e_{k,c,i}\, B_{k,c,i}(j),
  \label{eq:unitcap_lin}
\end{equation}
where $Y_j^{(0)}$ is the spectrum without mismatch, limited only by quantization, and $B_{k,c,i}(j)$ is the spectrum of the on/off pattern of unit $i$, held by slice $k$ as in \eqref{eq:spectrum}. Because $Y_j$ is a weighted sum of Gaussian errors, its real and imaginary parts are jointly Gaussian, centered on $Y_j^{(0)}$, with the $2\times2$ covariance matrix $\mathbf{C}_j = \sigma_u^2 \sum [\Re B; \Im B][\Re B; \Im B]^{\mathsf{T}}$. Rotating to the eigenvectors of $\mathbf{C}_j$, with eigenvalues $\lambda_1$ and $\lambda_2$, splits the spur power into two independent terms,
\begin{equation}
  S_j = \lambda_1 (\xi_1 + b_1)^2 + \lambda_2 (\xi_2 + b_2)^2,
  \qquad \xi_{1,2} \sim \mathcal{N}(0,1),
  \label{eq:unitcap_chi2}
\end{equation}
where $b_{1,2}$ are the components of $Y_j^{(0)}$ along the eigenvectors,
divided by $\sqrt{\lambda_{1,2}}$. This is a noncentral generalized $\chi^2$
distribution, and its cumulative distribution $F_j$ can be evaluated by a
one-dimensional numerical integral.

The dominant spur is the image at $-f_0$ caused by I/Q imbalance. If the I and Q arrays have slightly different total gains, a tone at $+f_0$ leaks an image at $-f_0$ whose amplitude is proportional to the gain difference, averaged over the $M$ slices. The phase of this image is fixed by the signal, and only its size and sign are random. One of the two terms in \eqref{eq:unitcap_chi2} therefore vanishes ($\lambda_2/\lambda_1 < 10^{-7}$), and the image power is the square of a single Gaussian variable, i.e., a scaled $\chi^2_1$ variable, as for the group $q = M/2$ in \Cref{sec:spur_stats}. Averaging over the slices makes its variance fall as $1/M$. Compared with the exponential distribution of a spur with random phase, the $\chi^2_1$ distribution has a long tail, which is why the $95\%$-yield budget below is about three times tighter than the median.

Spurious tones in different bins share the same unit errors; hence, they are correlated, but the distribution of the worst spur is bounded by $1-\sum_j (1-F_j) \le P(\max_j S_j \le t) \le \min_j F_j$, with the product $\prod_j F_j$ as the point estimate. Only the $64$ bins with the largest error variance or nominal level are kept; in a $2000$-trial Monte Carlo simulation of the full spectrum the worst bin never fell outside of this set. For a $55$\,dBc \ac{SFDR} the analysis gives a median budget of $\sigma_{\Delta C/C} \le 10.5\%$ ($M=4$, bounded by $10.2$ and $10.8\%$) and $15.0\%$ ($M=8$, bounded by $14.8$ and $15.3\%$) and a $95\%$-yield budget of $3.7\%$ and $5.3\%$, for $M=4$ and $M=8$, respectively. In this case, the bounds collapse because one spur is dominant. The median worst spur lies within $0.3$\,dB of the median I/Q image. Against the Monte Carlo, every point falls inside the $95\%$ confidence interval of the simulated quantile, with a largest deviation of $0.4$\,dB.

\subsection{In-Band SNDR Budget}
\label{sec:sndr}

The non-idealities in the prior subsections result in \ac{OOB} spurs. However, with a modulated signal, the in-band \ac{SNDR} is also important. The non-idealities are split into two classes. With static slice errors the array remains linear and periodic in $T_b$, so by \eqref{eq:spectrum} their errors land at $f + m f_b$, $m \neq 0$, outside the channel for $B < f_b$, apart from a common gain and delay that any receiver removes. With these removed, a Monte Carlo simulation at the median budgets of \Cref{sec:spur_stats} and \Cref{tab:clock}, with a $320$\,MHz \ac{OFDM}-like stimulus ($10$\,dB \ac{PAPR}), gives an in-band SNDR above $80$\,dB in $95\%$ of the trials. Inter-slice mismatch is therefore an \ac{OOB} emission problem only. Quantization, clock jitter and unit-capacitor mismatch, in contrast, place error inside the channel, and all of them add in power,
\begin{equation}
  \mathrm{SNDR}^{-1} = \mathrm{SNR}_q^{-1} + \mathrm{SNR}_j^{-1} + \mathrm{SNDR}_u^{-1}.
  \label{eq:sndr_sum}
\end{equation}

Each slice quantizes at $f_b$, so a fraction $B/f_b$ of its noise falls in band, and recombination adds $10\log_{10}M$ (\Cref{sec:resolution}). The resulting \ac{SNR} is:
\begin{equation}
  \mathrm{SNR}_q = \mathrm{SNR}_1 + 10\log_{10}\frac{f_\mathrm{RF}}{B},
  \label{eq:snr_q}
\end{equation}
where $\mathrm{SNR}_1$ is the Nyquist-band \ac{SNR} of a single slice. The in-band quantization noise is thus set by $f_\mathrm{RF}$, not by $M$: for the aforementioned signal and resolution, simulated with the stimulus peak at $0.95$ of full scale, $\mathrm{SNR}_1 = 45.5$\,dB and $\mathrm{SNR}_q = 61.4$\,dB for both variants.

\subsubsection{Clock Jitter}
Unlike skew, jitter changes from edge-to-edge; hence, it raises the noise floor rather than producing a spur. As in \Cref{sec:clocks}, the critical clock is $f_\mathrm{RF}$. If its edge in RF cycle $n$ at slice $k$ is displaced by $j_k[n]$, the held input word $h_k[n]$ of that slice is rotated by $\theta_k[n] = 2\pi f_\mathrm{RF}\,j_k[n]$, and to a first order the output error is $-j\sum_k \theta_k[n]\,h_k[n]$. The jitter is split into a common-mode (e.g., the source and the shared clock tree), with an rms $\sigma_\mathrm{cm}$ value, and an independent mode due to each slice's local buffer, with rms $\sigma_\mathrm{loc}$. The common part rotates the whole output, whereas the local parts add in power, and since $\sum_k |h_k|^2 \approx |y|^2/M$ for an oversampled signal they are averaged by $M$. For white jitter, a fraction $B/f_\mathrm{RF}$ of the error falls in band, giving:
\begin{equation}
  \mathrm{SNR}_j = \frac{f_\mathrm{RF}/B}{\left(2\pi f_\mathrm{RF}\right)^2
                   \left(\sigma_\mathrm{cm}^2 + \sigma_\mathrm{loc}^2/M\right)}.
  \label{eq:snr_jitter}
\end{equation}
For $\mathrm{SNR}_j = 55$\,dB this allows $\sigma_\mathrm{cm} = 140$\,fs, or $\sigma_\mathrm{loc} = 280$\,fs ($M=4$) and $396$\,fs ($M=8$), and \eqref{eq:snr_jitter} agrees with a Monte Carlo simulation to within $0.16$\,dB. The common term defines the phase-noise requirement of any \ac{RF-DAC} at $f_\mathrm{RF}$; when it is not white, its contribution is the phase noise integrated over the channel. At $f_\mathrm{RF} = 12.8$\,GHz, $\sigma_\mathrm{cm} = 140$\,fs corresponds to a white phase noise of $-140$\,dBc/Hz, i.e., to $-55$\,dBc integrated over the $320$\,MHz channel. Only the local term is specific to the architecture, and it is relaxed by $\sqrt{M}$. The $f_\mathrm{RF}$ jitter is required to be at femtosecond levels, whereas the static $f_\mathrm{RF}$ skew of \Cref{tab:clock} may reach $0.7$\,ps, because static errors leave only a common gain and delay in band, while jitter is white and falls in band directly. Because the data are retimed, the same edge also moves the data transition, but this adds only $M(B/2f_\mathrm{RF})^2/3$ of the carrier term, at most $-33.8$\,dB ($M=8$). The jitter of the phase clocks $\phi_k$ adds to their skew in \eqref{eq:slip_law} and is otherwise absorbed by the retiming.

\subsubsection{Unit-Capacitor Mismatch}
The unit-capacitor errors of \Cref{sec:unitcap} produce, besides the I/Q image, code-dependent distortion, and with a modulated signal both lie in band. Since \eqref{eq:unitcap_lin} is exactly linear in the unit errors, the in-band error after projecting out the common complex gain with $\boldsymbol{\Pi}_\perp$ is a quadratic form,
\begin{equation}
  \frac{P_u}{S} = \sigma_u^2\,\mathbf{e}^{\mathsf{T}}\mathbf{Q}_u\,\mathbf{e},
  \qquad
  \mathbf{Q}_u = \frac{1}{S}\,\Re\!\left\{\mathbf{B}^{\mathsf{H}}\boldsymbol{\Pi}_\perp\mathbf{B}\right\},
  \label{eq:unitcap_inband}
\end{equation}
where $S$ is the in-band signal power, $\mathbf{e}$ collects the $2MN_u$ unit errors $e_{k,c,i}$, and the columns of $\mathbf{B}$ are their in-band spectra $B_{k,c,i}(j)$. As for the spurs, this is a weighted sum of squared Gaussian variables, and its CDF follows from Imhof's integral~\cite{imhof1961quadratic}. Two properties set the budget. First, the mean error power, $\sigma_u^2\,\mathrm{tr}\,\mathbf{Q}_u$, falls as $1/M$, because the errors of different slices are independent and average out. Second, a single term dominates: the I/Q image carries $75\%$ of the error power, so the in-band error inherits the long tail of the image spur, and the $95\%$-yield budget is much tighter than the median one. For $\mathrm{SNDR}_u = 55$\,dB, the unit capacitors must match to $\sigma_{\Delta C/C} \le 5.3\%$ ($M=4$) and $7.5\%$ ($M=8$) at the median, and to $2.4\%$ and $3.4\%$ at $95\%$ yield. An independent time-domain Monte Carlo simulation of $2000$ trials reproduces the mean to within $1\%$ and both quantiles to within $6\%$ in power. These budgets are $1.6$ to $2$ times tighter than the \ac{SFDR} budgets of \Cref{sec:unitcap}: the \ac{SFDR} counts only the largest spur, whereas the in-band error collects all of the distortion and, relative to the signal, grows with the back-off. In-band SNDR, not \ac{SFDR}, therefore sets the size of the unit capacitors.

\begin{table}[t]
\caption{In-Band \ac{SNDR} Budget, for the $320$\,MHz \ac{OFDM} Stimulus}
\label{tab:sndr}
\centering
\begin{tabular}{lcc}
\hline
Term & $M=4$ & $M=8$ \\ \hline
Quantization $\mathrm{SNR}_q$ & $61.4$\,dB & $61.4$\,dB \\
Common $f_\mathrm{RF}$ jitter $\sigma_\mathrm{cm}^\dag$ & $140$\,fs & $140$\,fs \\
Local $f_\mathrm{RF}$ jitter $\sigma_\mathrm{loc}^\dag$ & $280$\,fs & $396$\,fs \\
$\sigma_{\Delta C/C}^\dag$, median / $95\%$ yield & $5.3$ / $2.4\%$ & $7.5$ / $3.4\%$ \\
Inter-slice mismatch$^\ast$ & $>80$\,dB & $>80$\,dB \\ \hline
Total SNDR, \eqref{eq:sndr_sum} & $51.5$\,dB & $51.5$\,dB \\ \hline
\multicolumn{3}{l}{\scriptsize $^\dag$Allowed value for $55$\,dB from that term alone.}\\
\multicolumn{3}{l}{\scriptsize $^\ast$At the median \ac{SFDR} budgets, $95\%$ of trials;}\\
\multicolumn{3}{l}{\scriptsize \phantom{$^\ast$}gain and delay removed.}
\end{tabular}
\vspace{-4mm}
\end{table}

\Cref{tab:sndr} summarizes the in-band budget. Like the \ac{SFDR} budgets, the jitter and unit-capacitor budgets scale by $10^{-(S-55)/20}$ for a different target $S$. With quantization at $61.4$\,dB and jitter and unit-capacitor mismatch each at $55$\,dB, \eqref{eq:sndr_sum} gives $51.5$\,dB.
\section{Optimizing the Number of Phases, M}
\label{sec:choosingM}

By using the yield laws derived in \Cref{sec:analysis} along with the timing overheads of a target process and standard cell library, we can determine how every requirement scales with $M$, before any implementation, by first establishing digital and analog design budgets and constraints.

\subsection{Digital Lower Bound}

The fastest digital clock runs at $f_\mathrm{RF}/M$ (\Cref{sec:general}), so each register stage has a cycle of $M\,T_\mathrm{RF}$, where $T_\mathrm{RF} = 1/f_\mathrm{RF}$. Out of this, every stage requires a fixed overhead that does not depend on $M$: the flip-flop clock-to-Q and setup times, $t_\mathrm{ov}$, and the jitter $t_j$ and skew $t_\mathrm{sk}$ of the clock. The time left for combinatorial logic is then:
\begin{equation}
  L(M) = M\,T_\mathrm{RF} - \left(t_\mathrm{ov} + t_j + t_\mathrm{sk}\right),
  \label{eq:stage_budget}
\end{equation}
which grows linearly with $M$. Pipelining cannot recover a negative $L$, because every added register requires the same overhead. Hence, once $t_\mathrm{ov} + t_j + t_\mathrm{sk} \ge T_\mathrm{RF}$, a direct-digital implementation ($M=1$) cannot close timing at any pipeline depth. Even a positive $L$ is not sufficient: the datapath contains stages that cannot be split further, and the longest of them, which no further pipelining can shorten, sets a minimum logic delay $D_\mathrm{min}$. Requiring $L(M) \ge D_\mathrm{min}$ gives the lower bound
\begin{equation}
  M \ge f_\mathrm{RF}\left(t_\mathrm{ov} + t_j + t_\mathrm{sk} + D_\mathrm{min}\right),
  \label{eq:Mmin}
\end{equation}
or, equivalently, for a fixed $M$ timing can be closed up to $f_\mathrm{RF,max} = M/(t_\mathrm{ov} + t_j + t_\mathrm{sk} + D_\mathrm{min})$. The bound depends only on the technology/library for the implementation and the longest indivisible stage, not on the filter length, and on $P$ only through whether $M < P$ (commutator crossing) or $M = P$ (adder tree).

\subsection{Cost of Raising M on Analog/RF}

The analog budgets of \Cref{sec:analysis} depend on $f_\mathrm{RF}$ only through the ratio $x = f_0 M/f_\mathrm{RF} = f_0/f_b$ (i.e., the position of the band edge relative to the nulls from the zero-order hold at $m f_b$). Hence, for a given signal bandwidth and $f_\mathrm{RF}$, the dimensionless budgets (gain, unit-capacitor, and phase-clock skew in degrees) are functions of $M$ alone; the $f_\mathrm{RF}$ skew and jitter budgets additionally scale as $1/f_\mathrm{RF}$. Increasing $M$ relaxes three of the specifications: the unit-capacitor budget grows as $\sqrt{M}$, because the I/Q image averages over $M$ independent slices, the local $f_\mathrm{RF}$ jitter budget also grows as $\sqrt{M}$, and the data-update skew budget increases, although retiming absorbs that skew in any case. The common jitter budget is independent of $M$, and all other budgets tighten, as shown here for a $55$\,dBc \ac{SFDR} at median yield with $f_0 = 160$\,MHz and $f_\mathrm{RF} = 12.8$\,GHz ($x = M/80$):
\begin{itemize}
  \item gain matching, from $7.7\%$ at $M=4$ to $1.8\%$ at $M=32$, because the images move up the hold's skirt as $x$ grows;
  \item the $f_\mathrm{RF}$ arrival, from $0.79$ to $0.23$\,ps;
  \item the multiphase generator, from $32^\circ$ to $2.4^\circ$ rms,
        because its slip window is fixed in time while its period shrinks
        in degrees.
\end{itemize}
Earlier analysis focused on specific cases of $M$. However, Monte Carlo simulations of other values of $M$ also support the yield laws. A $2\times10^4$-trial Monte Carlo at $M=2$ and $M=16$, with each mismatch set to its predicted median budget, lands within $0.13$\,dB of the $55$\,dBc target. The digital energy and area also grow with the number of slices, as each slice carries its own decoder and output drivers.

The unit-capacitor relaxation should not be extrapolated far. The
slice-differential spur sits $18$\,dB below the I/Q image at $M=4$, $11$\,dB at $M=8$, $6$\,dB at $M=16$, and reaches it at $M=32$. The image falls as $1/M$ while the differential spur grows with $M$. For $M \ge 32$, the differential spur sets the unit-capacitor budget.

\subsection{Design Rule}

The analysis above shows that, while increasing $M$ relaxes the digital timing, it tightens most of the analog requirements. Hence, the best design choice is the smallest $M$ that satisfies \eqref{eq:Mmin}, rounded up to the next divisor of $P$ (a power of two for $P = 8$),
\begin{equation}
  M^\ast = 2^{\left\lceil \log_2 f_\mathrm{RF}\left(t_\mathrm{ov} + t_j + t_\mathrm{sk} + D_\mathrm{min}\right)\right\rceil}.
  \label{eq:Mopt}
\end{equation}
As an example, consider the 28-nm library used in \Cref{sec:implementation}. In the slow-slow (SS) corner the flip-flop overhead is $t_\mathrm{ov} \approx 104$\,ps, and with budgets of $t_j = 50$\,ps and $t_\mathrm{sk} = 20$\,ps for clock jitter and skew, every stage requires $174$\,ps. The longest indivisible stage of the 4-phase datapath, the crossing from the FIR branches into the digital commutators, takes $D_\mathrm{min} \approx 98$\,ps. At $f_\mathrm{RF} = 12.8$\,GHz, \eqref{eq:Mmin} then requires $M \ge 3.5$, so $M^\ast = 4$. This choice holds for $7.4 < f_\mathrm{RF} \le 14.7$\,GHz, and $M=8$ is required up to $29.4$\,GHz. For $M = P$ there are no digital commutators, and $D_\mathrm{min}$ is set instead by the sub-filter adder tree (\Cref{sec:results}); the $29.4$\,GHz figure, like the $f_\mathrm{RF,max}$ column of \Cref{tab:choosingM}, therefore assumes the combiner $D_\mathrm{min}$ and is an estimate. Since the \ac{FR3} range targeted in \Cref{sec:polyphase} extends beyond $14.7$\,GHz, both variants are implemented in \Cref{sec:implementation}: the 4-phase variant as the choice at $12.8$\,GHz, and the 8-phase variant as the candidate for the upper \ac{FR3} range (signed off here at $16$\,GSps) and to quantify the digital cost of raising $M$ at a fixed $f_\mathrm{RF}$. \Cref{tab:choosingM} summarizes the resulting requirements for $M = 2$ to $32$ at $f_\mathrm{RF} = 12.8$\,GHz.

\begin{table}[t]
\caption{Requirements Versus $M$ for $55$\,dBc SFDR at $f_\mathrm{RF} = 12.8$\,GHz}
\label{tab:choosingM}
\centering
\begin{tabular}{lcccccc}
\hline
$M$ & $L(M)$ & $f_\mathrm{RF,max}$ & $\sigma_g$ & $\sigma_c$ & $\sigma_\phi$ & $\sigma_{\Delta C/C}$ \\
 & (ps) & (GHz) & (\%) & (ps) & ($^\circ$) & (\%) \\ \hline
$2$  & $-18$ & $7.4$  & $14.5$ & $1.26$ & $86$  & $7.4^\dag$  \\
$4$  & $138$ & $14.7$ & $7.7$  & $0.79$ & $32$  & $10.5$ \\
$8$  & $451$ & $29.4$ & $5.3$  & $0.67$ & $13$  & $15.0$ \\
$16$ & $1076$ & $58.9$ & $3.4$ & $0.44$ & $5.5$ & $21.2^\dag$ \\
$32$ & $2326$ & $118$ & $1.8$  & $0.23$ & $2.4$ & n/a$^\ddag$ \\ \hline
\multicolumn{7}{l}{\scriptsize Median-yield budgets. $\sigma_c$: $f_\mathrm{RF}$ arrival skew, retimed.}\\
\multicolumn{7}{l}{\scriptsize $\sigma_\phi$: phase-clock skew in slice update-clock degrees.}\\
\multicolumn{7}{l}{\scriptsize $^\dag$Extrapolated as $\sqrt{M}$.}\\
\multicolumn{7}{l}{\scriptsize $^\ddag$Set by the slice-differential spur, not by $\sqrt{M}$ scaling.}\\
\multicolumn{7}{l}{\scriptsize $P = M$ assumed for $M > 8$. $f_\mathrm{RF,max}$ uses the 4-phase combiner}\\
\multicolumn{7}{l}{\scriptsize $D_\mathrm{min} = 98$\,ps; for $M \ge P$, $D_\mathrm{min}$ is set by the sub-filter adder}\\
\multicolumn{7}{l}{\scriptsize tree and $f_\mathrm{RF,max}$ is lower (\Cref{sec:results}).}
\end{tabular}
\vspace{-4mm}
\end{table}

\section{Circuit Implementation and Layout}
\label{sec:implementation}

To validate the physical viability of the proposed hybrid architectures, both macros were implemented up to the layout level: the 4-phase variant, the choice of \eqref{eq:Mopt} at $12.8$\,GHz, and the 8-phase variant, to quantify the cost of raising $M$. The off-chip Farrow front-end resamples the modem stream to $f_\mathrm{RF}/32 = 400$\,MSps, and the macro performs the high-speed integer-rate interpolation, to be paired with a custom \ac{SCPA} acting as the RF-DAC for the passive analog recombination \cite{yuan2017_multiphase}. On the macro, an integer-rate $\times 4$ FIR interpolator first raises the sample rate from $f_\mathrm{RF}/32$ to the interleaved rate $f_\mathrm{RF}/8 = 1.6$\,GSps. Then, the polyphase FIR bank completes the $\times 8$ interpolation. For the 8-phase variant, the data is streamed out of the macro at $f_\mathrm{RF}/8 = 1.6$\,GSps, to be recombined on the RF-DAC. For the 4-phase variant, an additional digital commutator recombines the signals into 4-phases operating at $f_\mathrm{RF}/4 = 3.2$\,GSps, to be recombined on the RF-DAC. 

\begin{figure}[t]
  \centering
  \includegraphics[width=0.8\linewidth]{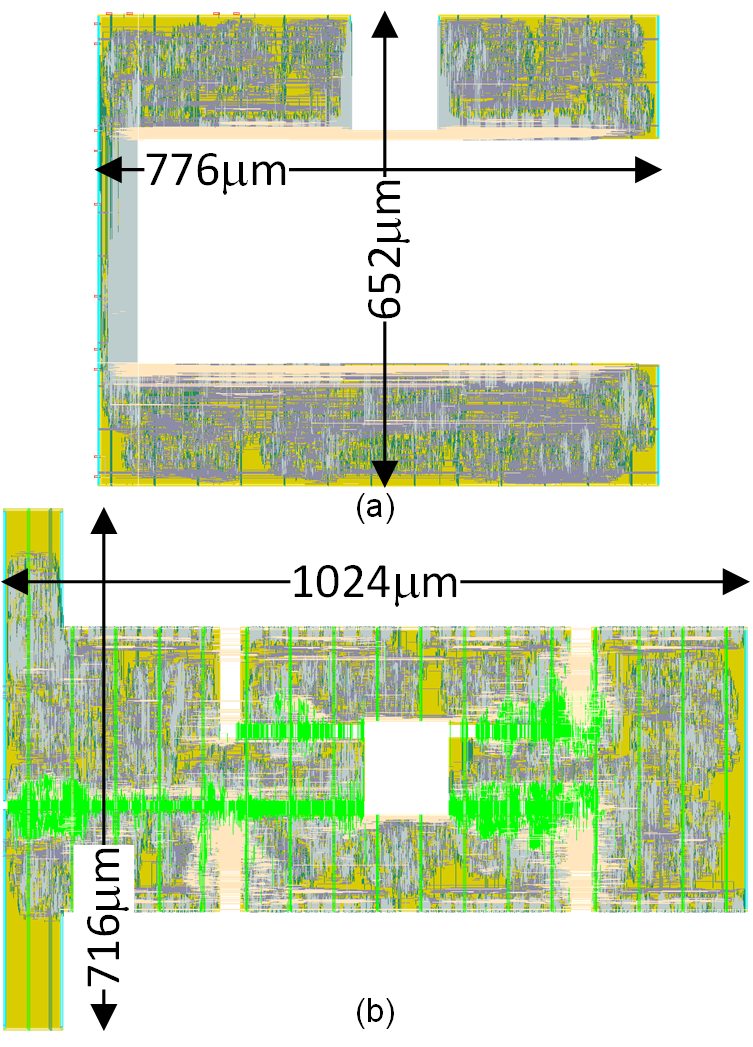}
  \caption{Placed-and-routed layouts of the digital macros in 28-nm CMOS. (a)~4-phase macro, $776\times652\,\mu$m ($0.51$\,mm$^2$). (b)~8-phase macro, $1024\times716\,\mu$m ($0.73$\,mm$^2$).}
  \label{fig:layouts}
\end{figure}

The digital polyphase FIR filter branches were generated as RTL from the Simulink model with HDL Coder and synthesized into a standard-cell gate-level netlist, optimizing the critical timing paths to ensure strict timing closure at $3.2$\,GHz ($f_\mathrm{RF}/4$) for the 4-phase variant and $1.6$\,GHz ($f_\mathrm{RF}/8$) for the 8-phase variant. The 8-phase macro was signed off at a $2$\,GHz slice update clock ($16$\,GSps effective output rate), $25\%$ above the nominal $1.6$\,GHz, with propagated clocks in the SS corner ($0.81$\,V, $125^\circ$C), using \ac{LVT} standard cells only. The 4-phase macro was signed off at $3.6$\,GHz ($14.4$\,GSps effective output rate), $12.5\%$ above the nominal $3.2$\,GHz, in the same corner, using \ac{ULVT} standard cells only; its polyphase FIR branches run at half this rate. \Cref{fig:layouts} shows the placed-and-routed layouts of both digital macros.

The physical routing of this multi-phase clock network presents a primary design challenge, particularly for the 8-phase variant, which distributes eight $45^\circ$-spaced clock phases. Because each slice retimes its data on $f_\mathrm{RF}$, skew in this network causes no spur as long as it stays inside the slip window of \eqref{eq:slip_law}, i.e., $22.5$\,ps rms at $M=8$ at the median (\Cref{tab:clock}). A symmetric, balanced H-tree routing topology was used across the layout, isolating the clock lines to minimize cross-coupling and differential phase skew~\cite{friedman2001_clock}. The $f_\mathrm{RF}$ distribution, which sets the binding sub-picosecond requirement, resides in the RF-DAC and is outside the scope of these macros.
\section{Post-Layout Validation and Results}
\label{sec:results}

Both macros were placed and routed in Cadence Innovus, and their parasitic resistances and capacitances (R+C) were extracted with Quantus. The implementation results of this section come from sign-off analysis of the extracted post-route netlists: static timing in Tempus, with propagated clocks, and power in Voltus, with switching activity from a gate-level Verilog simulation of the post-route netlist. To exercise the datapath with realistic activity, an 802.11be-like $320$\,MHz signal, chosen for its bandwidth and \ac{PAPR}, was processed through the behavioral MATLAB/Simulink model of the Farrow filter, and the resulting bitstreams drove the gate-level simulation. The output bit patterns of the post-route netlist, ideally reconstructed, were identical to those of the Simulink model over the full $40{,}799$-sample burst, confirming that both macros are bit-exact. This check is a zero-delay simulation and verifies function only; timing is covered by the static sign-off.

\begin{figure}[t]
    \centering
    \subfloat[]{\includegraphics[width=1\linewidth]{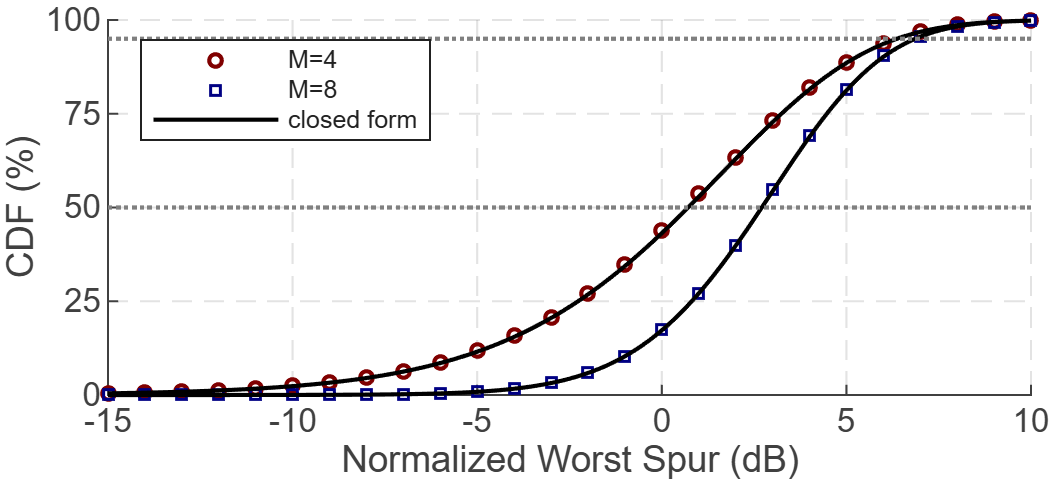}%
        \label{fig:sfdr_cdf_a}}
    \vspace{-4mm}
    \subfloat[]{\includegraphics[width=1\linewidth]{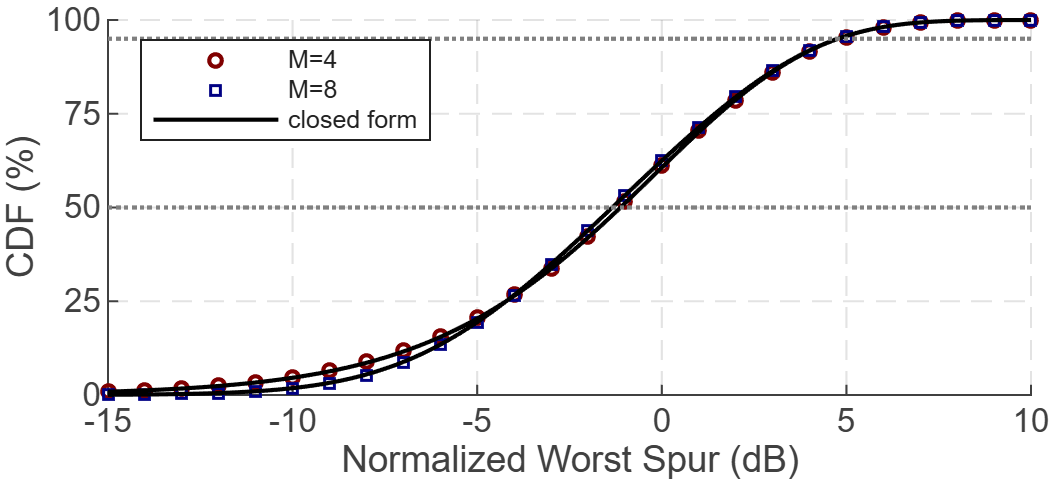}%
        \label{fig:sfdr_cdf_b}}
    \caption{CDF of the worst mismatch spur for (a)~timing skew and (b)~gain error, for a single tone at $f_0 = 160$\,MHz. The spur power is normalized to $\bar{\mu} = \max_q \mu_q$, the mean power of the strongest spur group across chips, from \eqref{eq:timing_law}--\eqref{eq:gain_law}. Markers: exact Monte Carlo of \eqref{eq:spectrum} over $2\times10^4$ trials; black lines: \eqref{eq:cdf_max}. Dotted lines mark the median and $95\%$ yield.}
    \label{fig:sfdr_cdf}
\end{figure}

The impact of the non-idealities of \Cref{sec:analysis} on the \ac{SFDR} is evaluated with a behavioral model of the 4- and 8-phase recombiners, with slice timing and gain errors that model clock-routing asymmetries and unit-capacitor mismatch. \Cref{fig:sfdr_cdf} verifies the yield law of \Cref{sec:spur_stats}: for each mechanism, the worst spur of $2\times10^4$ trials of the exact, non-linearized \eqref{eq:spectrum}, normalized to the largest group mean $\bar{\mu}$, follows \eqref{eq:cdf_max} for both variants. For gain error, the $M=4$ and $M=8$ curves nearly coincide, because the $q=1$ group dominates. For timing skew, all $M/2$ groups have the same mean, so the $M=8$ curve lies to the right, which is the yield cost of a larger $M$.

\begin{figure}[t]
    \centering
    \subfloat[]{\includegraphics[width=1\linewidth]{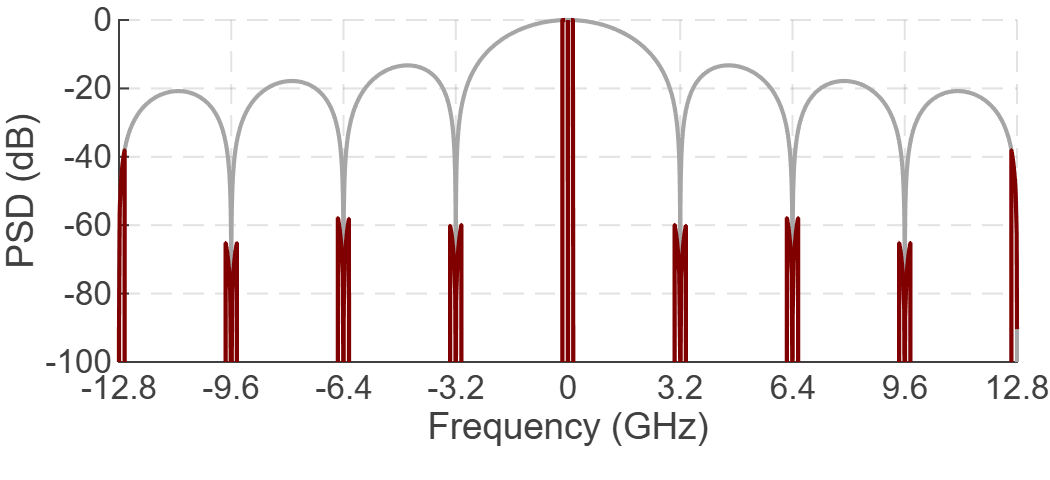}%
        \label{fig:spectrum_m_4}}
    \vspace{-4mm}
    \subfloat[]{\includegraphics[width=1\linewidth]{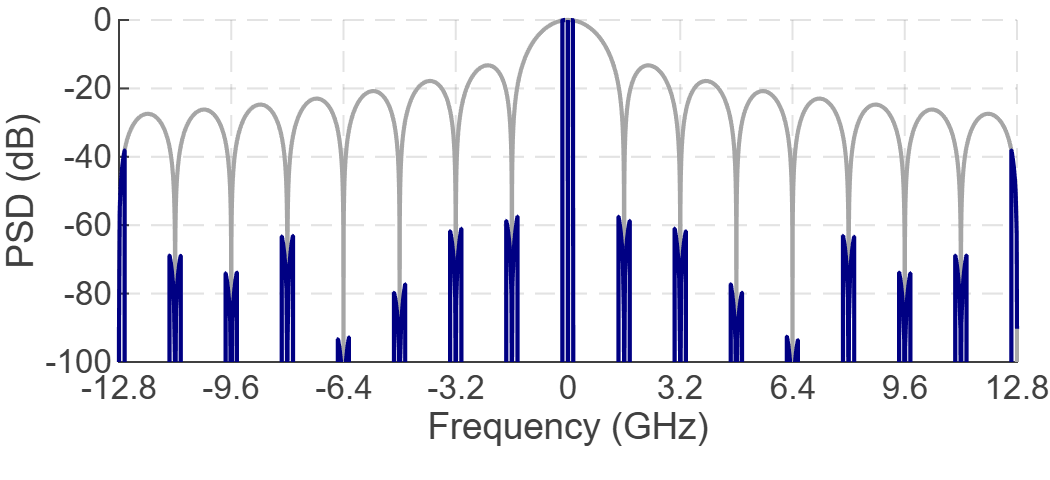}%
        \label{fig:spectrum_m_8}}
    \caption{Simulated output spectrum for a $320$\,MHz flat, random-phase multitone at $f_\mathrm{RF} = 12.8$\,GHz for (a)~$M=4$ and (b)~$M=8$, with slice timing skew $\sigma_\tau = 2$\,ps and gain error $\sigma_g = 3\%$; the median-SFDR draw of $1001$ Monte Carlo draws is shown. The PSD is normalized to the in-band PSD. Grey: hold envelope $\mathrm{sinc}(fT_b)$, with nulls at $m f_b$, $f_b = f_\mathrm{RF}/M$. Each mismatch image is the signal band shifted by $m f_b$, so it is centered on a null that notches its middle; with matched slices these images vanish. The bands at $\pm f_\mathrm{RF}$ are images of the recombined full-rate DAC, not mismatch products.}
    \label{fig:spectrum_mod}
\end{figure}

\Cref{fig:spectrum_mod} shows the output spectrum for a modulated stimulus with both errors applied. With matched slices, only the signal band and the full-rate images at $\pm f_\mathrm{RF}$ remain. Mismatch adds a copy of the band at every $m f_b$, about $60$\,dB below the in-band \ac{PSD} at its edges for the draw shown. Each copy straddles a hold null, so its center is notched and its band edges set its level, as assumed in \Cref{sec:spur_stats}.

For unit-capacitor mismatch, the output error splits exactly into a common-slice part, which by \eqref{eq:wide_zoh} is a single full-rate DAC carrying the errors averaged over the $M$ slices, and a differential-slice part, the only one that can break the polyphase cancellation. The worst resulting spur is the I/Q image of the common part; odd harmonics lie about $7$\,dB lower, and the differential part stays below the image up to $M=16$ (\Cref{sec:choosingM}). Averaging over the slices therefore gains about $10\log_{10}M$ over a single array, and since standard digital I/Q calibration removes the image, the budgets of \Cref{sec:unitcap} are conservative~\cite{anttila2008_frequency}.

Table~\ref{tab:comparison} provides a quantitative post-layout performance breakdown contrasting the conventional direct-digital approach against the implemented 4-phase and 8-phase hybrid variants.

\begin{table}[t]
\caption{Post-Layout Performance and Complexity Comparison}
\label{tab:comparison}
\centering
\resizebox{\columnwidth}{!}{
\begin{tabular}{lccc}
\hline
\textbf{Metric} & \textbf{\begin{tabular}[c]{@{}c@{}}Conventional\\ Direct-Digital\end{tabular}} & \textbf{\begin{tabular}[c]{@{}c@{}}4-Phase\\ Variant\end{tabular}} & \textbf{\begin{tabular}[c]{@{}c@{}}8-Phase\\ Variant\end{tabular}} \\ \hline
Nominal Output Rate & $12.8$\,GSps & $12.8$\,GSps & $12.8$\,GSps \\
Sign-off Rate (SS) & n/a & $14.4$\,GSps & $16$\,GSps \\
Nominal Digital Clock & $12.8$\,GHz & $3.2$\,GHz & $1.6$\,GHz \\
RF-DAC Slices ($M$) & 1 & 4 & 8 \\
Clock Phase Spacing & Single Phase & Quadrature ($90^\circ$) & Multi-Phase ($45^\circ$) \\
Timing Closure (SS) & Not achievable by \eqref{eq:Mmin} & Met at $3.6$\,GHz \P & Met at $2$\,GHz \S \\
Macro Energy / Sample (TT) $\|$ & --- & $27.2$\,pJ \P & $37.0$\,pJ \S \\
Power TT / SS / FF (mW) $\|$ & --- & $391$ / $414$ / $459$ & $592$ / $558$ / $710$ \\
Clock Tree, TT & --- & $30.0\%$ & $29.2\%$ \\
Leakage, TT / SS & --- & $6.3$ / $27.4\%$ & $2.6$ / $15.5\%$ \\
Digital Macro Area & --- & $0.51$\,mm$^2$ \P & $0.73$\,mm$^2$ \S \\
Routing Congestion & n/a & Medium & High \\ \hline
\multicolumn{4}{l}{\small TT: $0.9$\,V, $25^\circ$C; SS: $0.81$\,V, $125^\circ$C; FF: $0.99$\,V, $-40^\circ$C.} \\
\multicolumn{4}{l}{\small \S\ Implemented 8-phase macro, signed off and characterized at a $2$\,GHz} \\
\multicolumn{4}{l}{\small \phantom{\S\ }slice update clock ($16$\,GSps output).} \\
\multicolumn{4}{l}{\small \P\ Implemented 4-phase macro, signed off and characterized at a $3.6$\,GHz} \\
\multicolumn{4}{l}{\small \phantom{\P\ }slice update clock ($14.4$\,GSps output).} \\
\multicolumn{4}{l}{\small $\|$\ Whole digital macro, including the $\times 4$ interpolator.}
\end{tabular}
}
\end{table}

The benefit of a larger $M$ is timing feasibility rather than a reduction in arithmetic, as predicted by \eqref{eq:Mmin}. The 4-phase macro closes at $3.6$\,GHz ($14.4$\,GSps), $12.5\%$ above nominal and just below the $14.7$\,GHz bound of \Cref{sec:choosingM}; its critical path, the crossing from the half-rate FIR branches into the full-rate digital commutators, has $6$\,ps of slack. The 8-phase macro closes timing at $2$\,GHz, where its critical adder-tree stage meets setup with $17$\,ps of slack in the SS corner without additional pipelining.

Although the FIR arithmetic is the same for both variants, the implemented 8-phase macro carries about $1.9\times$ the datapath cell area of the 4-phase macro, excluding the $\times 4$ interpolator ($1.4\times$ in die area, $0.73$ versus $0.51$\,mm$^2$): it drives twice as many RF-DAC slices, each with its own thermometer decoder, and it must span the larger analog core of the 8-phase array, with correspondingly longer interconnect from the FIR branches to the lanes. Its energy per output sample, each macro at its own sign-off rate in the TT corner, is nevertheless only $1.4\times$ higher: $37.0$ versus $27.2$\,pJ for the whole macro (\Cref{tab:comparison}), or $34.5$ versus $24.5$\,pJ for the FIR datapath alone, excluding the $\times 4$ interpolator. It closes timing with LVT standard cells, whereas the 4-phase macro requires ULVT cells throughout. At their sign-off rates, the 8-phase variant thus buys timing margin for about $1.4\times$ the energy per sample, consistent with the 4-phase choice of \eqref{eq:Mopt}. A linear fit through the two macros gives about $17.4 + 2.45M$\,pJ per sample and $0.29 + 0.056M$\,mm$^2$; with two parameters for two points, mixed threshold-voltage flavors and sample rates, it indicates the trend rather than validating it.

Post-route power of both macros was annotated with switching activity from a zero-delay gate-level simulation of the post-route netlist driven by the 802.11be-like $320$\,MHz stimulus ($100\%$ of nets annotated; glitch power not included, so the reported energies are lower bounds), with each macro at its sign-off rate (\Cref{tab:comparison}). In the TT corner, the $\times 4$ interpolator consumes $40$\,mW ($6.8\%$, 8-phase) and $37.5$\,mW ($9.6\%$, 4-phase), and the clock tree about $30\%$ in both, so both macros are dominated by dynamic power and benefit directly from the reduced slice update clock. Leakage rises to $27.4\%$ of the 4-phase total at $125^\circ$C, because closing timing at $3.6$\,GHz required ULVT cells throughout.
\section{Conclusion}
\label{sec:conclusion}

This paper generalized polyphase interpolation for \acp{RF-DAC} by splitting the recombination between digital commutators and the \ac{RF-DAC} itself, which removes the full-rate output commutator and confines digital timing closure to $f_\mathrm{RF}/M$. This is particularly useful for wideband \acp{RF-DAC}, which cannot use high-order active reconstruction filters to suppress \ac{OOB} emissions. The slice array is equivalent to a single full-rate DAC with an $M$-times longer hold, and the yield laws, confirmed by Monte Carlo simulation, show that with retiming the binding requirement is sub-picosecond matching of the $f_\mathrm{RF}$ clock arrival across slices, and that the best choice is the smallest $M$ that closes digital timing.

Both implemented 28-nm macros close timing above their nominal rates in the SS corner: the 4-phase macro at $3.6$\,GHz ($14.4$\,GSps) and the 8-phase macro at $2$\,GHz ($16$\,GSps). The choice between them is a trade between timing feasibility and cost, not a power saving. The 8-phase variant consumes about $1.4\times$ the energy per output sample and die area of the 4-phase variant and needs eight $45^\circ$-spaced clock phases, but closes timing with LVT cells and has lower leakage at $125^\circ$C. The 4-phase variant therefore suits platforms that favor energy and only four clock phases, and the 8-phase variant suits those that need timing margin, high-temperature operation, or headroom to raise the output rate.

\renewcommand*{\bibfont}{\footnotesize}
\printbibliography

\end{document}